\documentclass[serv, nonblindrev]{informs-serv}

\OneAndAHalfSpacedXI
\usepackage[ruled,vlined]{algorithm2e}
\usepackage{float}
\usepackage{booktabs}
\usepackage{amssymb}
\usepackage{bbm}
\usepackage{color}
\usepackage{array}
\usepackage{comment}
\usepackage{thm-restate}
\usepackage{subcaption}
\usepackage{adjustbox}	
\usepackage[unicode]{hyperref}
\usepackage[title]{appendix}
\usepackage[style=authoryear,citestyle=authoryear,uniquename=false,maxcitenames=2,maxbibnames=9,natbib]{biblatex}
\usepackage[ruled,vlined]{algorithm2e}
\usepackage{float}
\usepackage{booktabs}
\usepackage{longtable}
\usepackage{multirow, color}

\TheoremsNumberedThrough     
\ECRepeatTheorems

\EquationsNumberedThrough  
\newcommand{\kibitz}[2]{\ifnum\Comments=0\textcolor{#1}{#2}\fi}

\begin{document}


\RUNTITLE{A Framework for the Joint Optimization of Assignment and Pricing}


\TITLE{A Framework for the Joint Optimization of Assignment and Pricing in Mobility-on-Demand Systems with Shared Rides}

\ARTICLEAUTHORS{%

\AUTHOR{Yang Liu\raise0.5ex\hbox{\textasteriskcentered}     }
\AFF{School of Civil and Environmental
Engineering, Cornell University, Ithaca, New York 14853, USA} 

\AUTHOR{Qi Luo\thanks{These authors contribute equally to this work. }\thanks{Corresponding author: Qi Luo, qluo2@clemson.edu}}
\AFF{Department of Industrial Engineering, \\ Clemson University, 277B Freeman Hall, Clemson, South Carolina, 29634, USA}

\AUTHOR{Raga Gopalakrishnan}
\AFF{Smith School of Business, Queen's University, Kingston, ON K7L 3N6, Canada}

\AUTHOR{Samitha Samaranayake }
\AFF{School of Civil and Environmental
Engineering, Cornell University, Ithaca, New York 14853, USA}
}

\date{}

\ABSTRACT{Mobility-on-Demand (MoD) systems have become a fixture in urban transportation networks, with the rapid growth of ride-hailing services such as Uber and Lyft. Ride-hailing is typically complemented with ridepooling options, which can reduce the negative externalities associated with ride-hailing services and increase the utilization of vehicles. Determining optimal policies for vehicle dispatching and pricing, two key components that enable MoD services, are challenging due to their massive scale and online nature. The challenge is amplified when the MoD platform offers exclusive (conventional ride-hailing) and shared services, and customers have the option to select between them. The pricing and dispatching problems are coupled because the realized demand depends on the quality of service (i.e., whom to share rides with) and the prices for each service type. We propose an integrated and computationally efficient method for solving the joint pricing and dispatching problem---both when the problem is solved one request at a time or in batches (a common strategy in the industry). The main results of this research include showing that: (i) the sequential pricing problem has a closed-form solution under a multinomial logit (MNL) choice model, and (ii) the batched pricing problem is jointly concave in the expected demand distributions. To account for the spatial evolution of supply and demand, we introduce so-called retrospective costs to retain a tractable framework. Our numerical experiments demonstrate how this framework yields significant profit increases using taxicab data in Manhattan, New York City, compared to dynamic dispatching with static pricing policies.}



\maketitle
 
\section{Introduction}
\label{sec:intro}

Mobility-on-Demand (MoD) systems have become a fixture in urban transportation networks, with the popularity of ride-hailing services operated by transport network companies (TNCs) such as Uber and Lyft. The popularity of such services has also stimulated the deployment of ridepooling \parencite{alonso2017demand} and on-demand microtransit \parencite{shaheen2020sharing}, and the study of the potential for shared automated vehicle systems \parencite{fagnant2014travel}.
The rapid growth of MoD systems, driven by the ubiquity of smartphones, electronic payments, and advances in vehicle dispatch algorithms, has changed the landscape for urban mobility and has the potential to reduce auto ownership. While such services in the form of ride-hailing (non-shared rides) are not likely to reduce the negative externalities associated with auto travel (e.g., pollution, congestion), and in-fact might even increase them~\parencite{FEHR}, high-capacity MoD applications (e.g., ridepooling, microtransit) have the potential to do so by improving the ratio between passenger miles traveled (PMT) and vehicle miles traveled (VMT)~\parencite{santi2014quantifying,alonso2017demand}. 
These studies provide insights into the potential benefits of high-capacity door-to-door mobility service as a more affordable, efficient, and sustainable alternative to the more prevalent ride-hailing services.

However, many questions remain regarding the availability and adoption of such services based on the ability of such systems to generate profits (relative to ride-hailing) and user preferences (e.g., the price convenience trade-off). 
Customers who appreciate convenience or privacy will seek exclusive services (i.e., not sharing rides with others). 
In contrast, others may prefer to use low-cost services as long as delays due to detours are acceptable.
In short, the proportion of customers opting for the shared option in an MoD platform depends on (a)~the perception of quality of service (QoS) from exclusive and shared rides and (b)~their prices. 
Therefore, MoD platforms need to optimize their services to maximize gross profits subject to operational constraints and commuter choice---requiring the development optimal vehicle dispatching and trip pricing mechanisms across the services that they provide. 
For example, in the case of an operator that provides ride-hailing and ridepooling options, the goal is to optimize both vehicle dispatch and pricing for both services, which leads to a complex, coupled optimization problem.  
\smallskip


\noindent \textbf{Joint vehicle dispatching and pricing problem (JVDPP).} \quad 
Approaches for optimal vehicle dispatching and trip pricing have been extensively investigated in the literature.
Vehicle dispatching involves matching available vehicles with pending requests and directing empty vehicles towards locations likely to have future requests. 
Efficient dispatching algorithms can reduce wait times and increase the overall throughput \parencite{alonso2017demand}. 
Pricing is an equally important instrument in the MoD platform because it controls the demand for services and profits. In MoD systems, pricing is used to balance supply and demand temporarily \parencite{ke2020pricing}, via what is commonly known as surge pricing, as well as in the long run \parencite{bimpikis2019spatial}, via spatial pricing.
For example, a pricing policy can simultaneously control the demand side by adjusting trip fares and control the supply side by redistributing fare splits \parencite{he2018pricing, zha2018geometric}. 


In our context, where an operator provides multiple service types, the platform must optimize vehicle dispatching and pricing jointly for the mixed service types and trip requests, which we refer to as the JVDPP.
This problem is non-trivial because the induced demand depends on both of these operational policies. 
A menu of exclusive and shared services with corresponding prices and estimated journey times is offered to each customer accessing the platform with a ride request. 
The rider is then assumed to choose the utility-maximizing option, while the MoD platform is expected to ensure that all customers who have ride options are served.
The discrete trip assignment decision and continuous pricing decision are coupled in JVDPP, leading to an optimization problem that is challenging to solve.

\subsection{Preliminaries and Related Work}
This section summarizes the growing literature on vehicle dispatching and pricing, and highlights how this work integrates these two streams into a cohesive joint optimization framework for MoD platforms. 
Table \ref{tab:1} summarizes the methods and findings in recent studies that focus on either topic or the joint problem.
We refer the reader to \citet{wang2019ridesourcing} 
for a comprehensive review of related studies.

\begin{table}[!htb]
	\centering
	\caption{Summary of related work} \label{tab:1}
	\begin{tabular}{l|l|l|l}
		\toprule 
		Topic  
		& Setting & Methods  & Reference \\ \hline
		\multirow{4}{*}{ 
			\begin{tabular}[c]{@{}l@{}}
				Dispatching, \\
				 assignment\\ 
				and rebalancing
		\end{tabular}     }                            
		& \multirow{2}{*}{Single-ride } & Sequential dispatch & \begin{tabular}[c]{@{}l@{}}
			\citet{zhang2017taxi} \\ 
			\citet{lei2020efficient} 
		\end{tabular}     \\ \cline{3-4} 
		&                   & Batched dispatch &   
		\begin{tabular}[c]{@{}l@{}}
			\citet{jintao2020learning} \\
			\citet{qin2020optimal}  \end{tabular}    
		\\ \cline{2-4} 
		& \multirow{2}{*}{Ridepooling } & 	\begin{tabular}[c]{@{}l@{}}   Dynamic \\  programming \end{tabular}     &  \begin{tabular}[c]{@{}l@{}} 
			\citet{yu2019integrated} \\
			\citet{shah2020neural} \\
		\end{tabular}    \\ \cline{3-4}
		&  & \begin{tabular}[c]{@{}l@{}} Combinatorial/ \\ 
			heuristic  \end{tabular}      & 	\begin{tabular}[c]{@{}l@{}} \cite{alonso2017demand} \\ 
			\citet{liu2018framework} \\
			\citet{luo2021efficient} \\
			\citet{simonetto2019real} \\
			\citet{lowalekar2021zone}
		\end{tabular}   
		\\ \cline{2-4}  \hline
		\multicolumn{1}{l|}{\multirow{3}{*}{Pricing}}                                              & \multicolumn{1}{l|}{\multirow{2}{*}{Single-ride  }} & \multicolumn{1}{l|}{ Aggregate model } & \multicolumn{1}{l}{ \begin{tabular}[c]{@{}l@{}} 
				\citet{castillo2017surge} \\ 
				\citet{he2018pricing} \\
				\citet{zha2018geometric}  \\
				\citet{guda2019your}
		\end{tabular}  }   \\ \cline{3-4} 
		\multicolumn{1}{l|}{}                                                               & \multicolumn{1}{l|}{}                  & \multicolumn{1}{l|}{ 	\begin{tabular}[c]{@{}l@{}}   Dynamic \\  programming \end{tabular}     } & \multicolumn{1}{l}{ \begin{tabular}[c]{@{}l@{}} 
				\citet{sayarshad2018scalable} 
		\end{tabular}  }   \\ \cline{2-4} 
		\multicolumn{1}{l|}{}                                                               & \multicolumn{1}{l|}{Ridepooling }                  & \multicolumn{1}{l|}{ Network modeling } & \multicolumn{1}{l}{ 
			\begin{tabular}[c]{@{}l@{}} 
				\citet{zhang2019pool} \\
				\citet{ke2020pricing}
			\end{tabular} 
			
		}   \\ \hline
		
		\multirow{2}{*}{\begin{tabular}[c]{@{}l@{}} Joint vehicle dispatch \\ and pricing  \end{tabular}} 
		&   Single-ride   & 	\begin{tabular}[c]{@{}l@{}}   Dynamic \\  programming \end{tabular}    &  \begin{tabular}[c]{@{}l@{}} 
			\citet{chen2020dynamic}  \\ 
			\citet{banerjee2017pricing} \\ 
			
		 \cite{ozkan2020dynamic} \end{tabular}      \\ \cline{2-4} 
		&   \begin{tabular}[c]{@{}l@{}} Mixed pooling\\    fleet   \end{tabular}      &  \begin{tabular}[c]{@{}l@{}} Combinatorial/ \\ 
			heuristic  \end{tabular}    & 
				\begin{tabular}[c]{@{}l@{}} 	
				\citet{yan2020dynamic} \\ 
			\textbf{This work}   \end{tabular}  
		\\
		\bottomrule
	\end{tabular}

\end{table}

\noindent \textbf{Vehicle dispatch methods.} \quad 
Two standard vehicle dispatch methods are sequential (continuous/instant) dispatch and batched dispatch. 
The sequential dispatch method assigns each request to nearby available vehicles upon their arrival \parencite{zhang2017taxi, lei2020efficient}.
This method is 
simple to implement and has modest performance \parencite{zha2018geometric}. 
A batched dispatch method collects requests during a short interval and solves a global assignment problem at the end of each interval. 
Putting trip requests in batches can potentially find a better matching with available vehicles or shareable rides compared to the sequential method \parencite{yan2020dynamic, jintao2020learning, qin2020optimal}.

The new operational challenge of vehicle dispatch for ridepooling services is how to combine multiple trip requests going in a similar direction into a single ride by solving the \emph{trip-vehicle assignment} problem.
Most models are limited to solve ridepooling with two customers per vehicle \parencite{santi2014quantifying, sundt2021heuristics}.
The computational challenge of solving the trip-vehicle assignment increases substantially when the vehicle capacity exceeds two.
However, state of the art algorithms for ridepooling can now solve large-scale problem instances with larger vehicle capacities in real-time. In one of the first such works, \citet{alonso2017demand} proposed an anytime-optimal algorithm that dynamically assigns multiple requests to vehicles by extending the idea of the shareability graph from \citet{santi2014quantifying} and exploiting the sparsity of the number of feasible trip configurations in practical problem instances. There is also a growing literature on proactive rebalancing/routing of idle/partially occupied vehicles to meet estimated future demand using historical data \parencite{bent2007waiting, wen2017rebalancing, spieser2016shared, liu2020proactive, fielbaum2021demand} and improving the computational efficiency \parencite{simonetto2019real, luo2021efficient}. 
All of these approaches are myopic in the sense that they do not consider future demands. The non-myopic problem formulations are significantly harder as they are high-dimensional stochastic optimization problems in a setting where the myopic problem is already very challenging. \citet{yu2019integrated} proposed an approximate dynamic programming approach with a spatial-and-temporal decomposition heuristic for improving the computational efficiency, but even this approach has limited scalability. To counter this challenge, \citet{shah2020neural} utilize a neural network based approach to learn the value function, which allows for solving larger problem instances.   
Deep reinforcement learning has also recently been adopted in industry \parencite{tang2021value,qin2020ride}. 



\noindent \textbf{Pricing methods.} \quad  Pricing can improve the revenue management of MoD systems by smoothing out the supply and demand imbalance temporarily through surge prices or permanently through optimal prices at steady states. There is a large and growing literature on this topic. \citet{banerjee2017pricing} modeled the MoD system as a continuous-time Markov chain where the state is the number of vehicles at each vertex and solved the optimal prices for each Origin-Destination (O-D) pair. In perhaps the most well known (currently) analysis of this pricing strategies for ride-hailing systems,   
\citet{castillo2017surge} proposed an aggregate model for relieving the so-called Wild Goose Chase (WGC) phenomenon in ride-hailing systems, using data-driven optimization. 
\citet{qiu2018dynamic} studied the pricing problem as a dynamic program that focuses on the temporal discrepancy of the problem. 
Other studies addressed problems such as demand-supply imbalance, cost split, and driver incentives \parencite{he2018pricing,guda2019your,zhang2019pool,ke2020pricing, bimpikis2019spatial, ozkan2020dynamic}.

\noindent \textbf{Research opportunities.} \quad 
The above works study the vehicle dispatch and pricing problems separately, and ignore the following operational issues which have not been fully addressed in the literature.
In many of the pricing models, trip fares are computed from aggregate models of operations, and therefore may not be consistent with what happens in practice. 
Furthermore, even though platforms typically provide shared and exclusive services simultaneously, prior studies typically optimize the systems independently. They do not determine the prices and vehicles dispatch strategies for both options in tandem, even though both options are available to each potential customer and compete with each other.
The network effect, which reflects the connection between the elasticity of supply and demand, has not been well-addressed in the literature, especially regarding strategic drivers' relocation decisions and customers' choices between different service options.
In this respedt, the work of \citet{yan2020dynamic} is the closest to ours. 
They studied the JVDPP in the conventional ride-hailing setting (i.e., exclusive services) or ridepooling under a simple time-window-based policy. 
A queueing-based heuristic determined the surge prices and matching time intervals at the market equilibrium.
In contrast, we consider an MoD platform with mixed fleets where pricing and dispatching decisions are made at the trip level, and develop MIP formulation that can compute the optimal vehicle dispatching and prices in real-time.




\subsection{Overview and Main Contributions}

We consider a transportation system with three travel modes that coexist---exclusive MoD service (e.g., UberX), shared MoD service (e.g., Uberpool), and an outside alternative (e.g., taxi service).
An MoD platform operates a mixed fleet of two services to satisfy different consumer groups. 
\emph{Endogenous demand} is considered throughout our analysis.
Particularly, customers will make the mode choice based on the prices and QoS, which follows a multinomial logit (MNL) model \parencite{liu2018framework, qiu2018dynamic}. 

The main goals of this work are to develop the following:
\begin{enumerate}
    \item A sequential pricing and dispatch (SPD) framework that assigns each trip request to an available vehicle and simultaneously determines the optimal prices.
    \item A batched pricing and dispatching (BPD) framework that determines globally optimal (for all requests in the batch) trip assignment policies and prices over fixed intervals.
\end{enumerate}

\noindent The corresponding technical contributions include: 
\begin{enumerate}
	\item Developing an exact approach that determines the optimal prices and vehicles to dispatch in real-time for exclusive and shared MoD services by harnessing the power of multi-product pricing.

 	\item 
	Proving that the objective of the JVDPP problem is jointly concave for the special case that at most two customers can share rides, and deriving analytical forms for optimal prices.
	
	\item Establishing a non-myopic mixed integer programming (MIP) framework by introducing a separable cost structure consisting of instantaneous costs and forward-looking control-dependent costs (termed ``retrospective costs''). 
	
\end{enumerate}

The introduction of retrospective costs is a simple and robust approach for decoupling the endogenous supply and demand constraints---by considering the duration a vehicle is blocked from accepting future requests after being assigned.  
A dispatching policy that considers retrospective costs allows for more tractably optimizing supply configurations based on future demands.
Since each customer may choose between two service types, we develop an \textit{overbooking} policy in BPD to ensure that vehicles are available for both options over the planning horizon.
These techniques can substantially reduce the computational challenge in other formulations (e.g., the dynamic programming formulation in \citet{yu2019integrated}). 




The remainder of the article is organized as follows:
Section~\ref{c4sec: problem} describes the background of the JVDPP and motivates a flexible and computationally efficient optimization framework to solve it.
Section~\ref{c4sec: experiments} discusses the experimental results using the taxi data in Manhattan, New York City (NYC).
The conclusions are drawn in Section~\ref{sec:conclusion}.
The notation used in this work is summarized in Table \ref{table:app1} in Appendix \ref{sec:app1}.

\section{Problem Description and Model Formulation}\label{c4sec: problem}

\subsection{JVDPP on MoD Platforms with Mixed Fleets}

We consider an  
MoD platform providing two types of services: exclusive service (indicated by subscript ``$e$'') and shared service (indicated by subscript ``$s$''). 
Exclusive service is the conventional ride-hailing service where each customer takes a private ride.
Shared service corresponds to a ridepooling option where customers agree to share rides with others on a similar route and pay reduced fares. 
As the exclusive service is more desirable (e.g., is faster, more private, and more reliable) and costly to operate, it is usually priced higher than the shared option. The MoD platform offers both premium and budget options to cater to various types of potential customers and these two options provide commuters service-specific prices and expected wait times.

\subsubsection{Overview of SPD and BPD.   }
\begin{figure}[!htb]
	\centering
	\includegraphics[width = 0.95\textwidth]{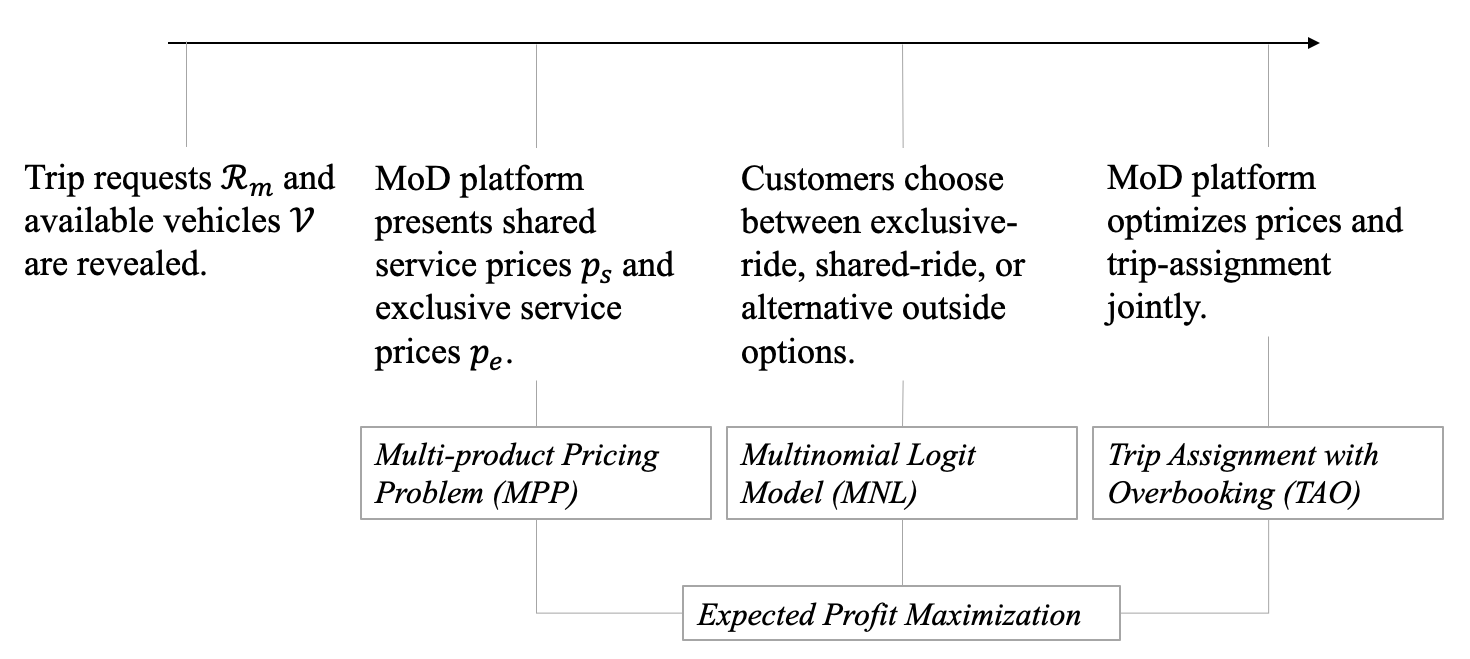}
	\caption{Road-map for implementing the BPD policy}
	\label{fig:1}
\end{figure}

SPD and BPD are two common operational frameworks adopted by TNCs. 
SPD assigns trip requests to available vehicles upon arrival. 
BPD allows new trip requests and vehicles in each region to wait for a fixed time interval to be matched so the platform can determine the dispatch plans and prices in batches.
The sequence of interactions between the arriving customers and the MoD platform is described in Figure \ref{fig:1}. 

Both frameworks have their respective merits and the overall performance depends on the market conditions (i.e., the supply and demand in the transportation network) \parencite{qin2020optimal}.
Since the prices and the QoS of exclusive and shared services affect how each customer chooses between them (possibly with outside alternatives also present), the pricing and dispatching problems are tightly coupled.
This research will simplify the analysis under the following assumptions: (a)~customers make irrevocable decisions after being assigned to rides or choosing alternative outside options, and (b)~drivers specify their preferred service type before the assignment.  
These assumptions are commonly imposed in the literature \parencite{qiu2018dynamic}.
Considering more sophisticated driver and customer behavior (e.g., ride cancellation) is possible in our general framework. 

\subsubsection{Preliminaries for trip-vehicle assignment (TVA).} 
\label{sec212}
Vehicle dispatching in SPD can be easily addressed by rule-based assignment (e.g., customers are matched with nearest available vehicles in \citet{zhang2019pool}) or optimal assignment based on an objective (e.g., minimizing total system travel time or passenger inconvenience in \citet{ota2016stars}).   
Since each assignment is made instantaneously, there is little room for improvement from considering more sophisticated online matching algorithms \parencite{ashlagi2018maximum}.

However, the problem is more computationally challenging in BPD, especially for high-capacity vehicle settings. 
One of the most popular approaches for high-capacity ridepooling is based on the work of \citet{alonso2017demand}, which develops an efficient TVA formulation for vehicle dispatching in BPD.
Our work builds on this general formulation, which we briefly introduce here for completeness. 
TVA describes the matchable relationship between the system's \textit{supply} (available vehicles) and \textit{demand} (pending trip requests) on a so-called shareablity graph \parencite{santi2014quantifying} (See Figure \ref{fig2}). 
\citet{alonso2017demand} proposed a multi-step procedure from constructing this graph to solving the optimal assignment---leading to a computationally efficient approach for solving vehicle dispatch problem in high-capacity MoD settings.



Whenever a new request $r$ arrives in the system, a set of criteria is applied to check whether each vehicle $v$ can be matched with the request and whether any existing request $r' \in \mathcal{R}$ can be matched with $r$. This allows the creation of a request and vehicle compatibility graph known as the shareability graph. 
Requests are assignable to a vehicle if the following conditions are satisfied:
\begin{enumerate}
	\item The wait time of each customer must be below the maximum wait time $\Omega$.
	\item The total delay~\footnote{The total delay is the difference between the trip's actual arrival time and the earliest possible arrival time.} must not exceed the maximum delay $\Delta$. 
	\item The number of customers in the vehicle has to be smaller than or equal to the capacity at any time.
\end{enumerate}

The shareability graph is then use to compute all feasible vehicle request pairings are determined at a predefined batching window (e.g., 30 seconds), utilizing the fact that any feasible grouping (trip) must form a clique in the shareability graph (a necessary but not sufficient condition for feasibility). One the feasible trips are computed, an ILP is solved to determine the optimal trip-vehicle assignment.    
This procedure also rebalances the vehicle fleet by routing empty vehicles to locations with potential for future demand at the end of each batch.
Since the main contribution of our work is to integrate pricing with the TVA formulation in \citet{alonso2017demand}, we focus on the problem of pricing at the request level in the remainder of this paper.

\subsubsection{Profit-maximizing pricing for mixed fleets. }


Trip fares for the exclusive service $p_e$ and shared service $p_s$ are presented to each customer entering the platform.
Let $P_e$, $P_s$, and $P_o$ denote the probability of choosing the exclusive services, the shared service, and some alternative outside options, such that $P_e + P_s + P_o = 1$. 
A set of trip requests $\mathcal{R}$ are revealed one by one in SPD or in batches in BPD.
The expected profit obtained by serving these demands $\mathcal{R}$ under a fixed pricing policy $\pi_p$ and an assignment policy $\pi_{a}$ is given by:
\begin{align} \label{eq1}
  	\mathbb{E}_{\pi_p, \pi_{a} } \Big[ \sum_{r \in \mathcal{R} } \Phi_r \Big]  = \sum_{r \in \mathcal{R}_e} P_e (p_e^r - c_e^r ) + \sum_{r \in \mathcal{R}_s }   P_s (p_s^r - c_s^r), 
\end{align}
where $\Phi_r$ is the profit from request $r\in \mathcal{R}$.
The MoD platform's goal is to search for optimal policies that maximize the expected profit.
More specifically, policy $\pi_p$ determines prices for exclusive service $p_e$ and shared service $p_s$ for each request $r\in \mathcal{R}$; policy $\pi_{a}$ determines the vehicle assignment process, either in a sequential or batched manner, for matching requests and available vehicles.   
Costs for the exclusive option are $c_e$ and for the shared option are $c_s$.  
Since the prices and assignments are determined at the request level for each $r \in \mathcal{R}$,  the trip request set $\mathcal{R}$ is split into demand served by exclusive service $\mathcal{R}_e$ and by shared service $\mathcal{R}_s$.   
Due to the existence of outside alternatives, $\mathcal{R}_e \cup \mathcal{R}_s \subset \mathcal{R}$.
Throughout, we omit the superscript $r$ whenever it is clear from the context. 
 
Direct calculation of the expected profit is not practical in JVDPP due to the challenge of estimating the continuation value, which is the expected future profit under the current pricing and dispatching policies.
For example, a dynamic program formulation models all evolutionary dynamics of supply and demand and seeks non-myopic policies, which is computationally intractable due to the ``curse of dimensionality''. 
This work seeks to develop a structured model that \emph{approximates} the system dynamics and customers' behavior under various pricing strategies and finds near-optimal policies in a more tractable manner. 


The proposed method assumes a centralized dispatch system where drivers follow the MoD platform's guidance strictly, and the derived method is appropriate for applications such as shared automated vehicles \parencite{fagnant2014travel}, high-capacity on-demand transit \parencite{alonso2017demand}, and ride-hailing platforms with permanent employees \parencite{dong2020optimal}.
The following sections discuss how to efficiently seek the optimal SPD and BPD policies with mild assumptions. 

\subsection{Pricing in SPD Framework} \label{sec: seq}
This section described how to set prices under the SPD framework. 
SPD is simple to implement and short matching times can strengthen customers' adherence to the current MoD platform in the face of competition.
The main challenge is the estimation of the expected profit obtained from serving requests in Equation~\eqref{eq1}, which includes the option of rejecting them to keep the vehicle open for future demand (e.g., if the request is sending the vehicle to a low demand region at a high demand time period). 
Denote feasible vehicles for a request by $\mathcal{V}_e$ for the exclusive service and  $\mathcal{V}_s$ for the shared service, respectively.
The SPD framework handles an arbitrarily arriving request $r$ as follows: 
\begin{enumerate}
	
	\item For vehicles for exclusive service $v_e \in \mathcal{V}_e$ and those for shared service $v_s \in \mathcal{V}_s$, the platform computes the optimal prices $p_e^*$ and $p_s^*$ 
	and picks a vehicle from each type as 
	$v_e \in \arg \min_{v_e' \in \mathcal{V}_e} \{ c_e(v_e', r) \}$ and  $v_s \in \arg \min_{v_s' \in \mathcal{V}_s}\{ c_s(v_s', r)\}$, respectively. The cost $c_e(v_e, r)$ (respectively, $c_s(v_s, r)$) measures the operational and retrospective costs associated with assigning request $r$ to vehicle $v_e$ (respectively, $v_s$). 
	
	\item The platform presents a menu of the best prices of each mode along with an outside alternative. 
	
	\item If the customer declines the MoD service, the procedure terminates; otherwise, the request is assigned to the corresponding vehicle.
	
\end{enumerate}

The key step in obtaining an explicit form for the profit function \eqref{eq1} is separating each cost into the \emph{operational} cost and the \emph{retrospective} cost. 
Operational costs consider all costs incurred in picking up customers and searching for rides (e.g., fuel, personnel). 
Retrospective costs include penalties for lost demands. 
The remainder of this section focuses on finding all feasible vehicles (Step 1) and computing optimal prices $p_e^*, p_s^*$ for each trip request (Step 2).  

Given that both types of vehicles are available, the pricing policy $\pi_p$ aims to compute two prices $p_e$ and $p_s$ to maximize the expected profit in Equation~\eqref{eq1}:
\begin{equation}\label{c4equ: seqobj}
	(p_e^*, p_s^*)= \arg \max \mathbb{E}_{\pi_p, \pi_{a}}  [\Phi_r ] =  \arg \max \{  P_e (p_{e} - c_e  ) + P_s  (p_{s} - c_s  )  \}.
\end{equation}

As the pricing decision is \emph{non-myopic}, the key challenge is to estimate the value function (i.e., the continuation value of making a pricing and assignment decision for $r$). 
Since Equation~\eqref{c4equ: seqobj} addresses that the mode choice parameters $P_e, P_s$ and the costs $c_e, c_s$ are both dependent on the pricing policy $\pi_p$, we can find analytical forms for $\Phi_r$ and derive a heuristic-based pricing policy detailed below.


\subsubsection{MNL model for service type choice.} 
The expected profit per request depends on each customer's mode choice parametrized by $P_e$ and $P_s$.
The MNL model is used for customers' choice between two MoD service types and an outside alternative. 
This model has been widely used in the revenue management and transportation literature \parencite{li2011pricing, he2018pricing, qiu2018dynamic}. 
Customers choose from a menu of modes and their corresponding prices, an option that maximizes their expected utilities.
Let $U_e$, $U_s$, and $U_o$ denote the respective utilities associated with three travel modes (exclusive MoD service, shared MoD service, and outside alternative). 
Their utilities for each travel mode $m \in \{e, s, o\}$ are given by:
\begin{equation}\label{equ:u}
	U_{m} = \beta_{p} \, p_{m} + \beta_{w} \, w_{m} + \beta_{t} \, t_{m}, 
\end{equation}
where $p_{m}$, $w_{m}$, and $t_{m}$ are the trip price, estimated wait time, and estimated travel time for choosing mode $m$. 
All these terms are computed after an arbitrary trip request $r$ is revealed. 
$\beta_{p}$, $\beta_{w}$, and $\beta_{t}$ 
are the corresponding coefficients estimated from historical data \parencite{liu2018framework}.  
Note that $\pi_p$ affects the first term in Equation~\eqref{equ:u} and $\pi_{a}$ affects the other terms.
The current setting considers homogeneous customers throughout the analysis for simplicity, but this assumption can be relaxed if there are additional data sources.  

The predicted probabilities of choosing travel mode $m$ are: 
\begin{equation}
	P_m = \frac{e^{U_{m}}}{e^{U_{e}} + e^{U_{s}} + e^{U_{o}}}\quad\quad \forall m \in \{e, s, o\},
\end{equation}
where the utilities of choosing modes $m \in \{e,s,o\}$ can be rewritten as: 
\begin{align*}
	U_m = U_m^{\pi_p} + U_m^{\pi_{a}} =  \beta_{p} \, p_m +  U_m^{\pi_{a}}.
\end{align*}

The following analysis focuses on  how to compute prices that approximately optimize Equation~\eqref{c4equ: seqobj} 
in the SPD framework, which assumes that
profits earned from serving trip requests in $\mathcal{R}$ are drawn from a known distribution (e.g., the empirical distribution).


\subsubsection{Operational and retrospective costs.}
The costs $c_e$ and $c_s$ depend on the pricing policy at the request level, and consists of two parts: the operational cost $c_{ \text{operational} } $ and the retrospective cost $c_{ \text{retrospective} }$. 
The operational costs include fuel, maintenance, and fixed tiers that may vary for different modes.
For exclusive service, the fixed operational cost $c_{e,  \text{operational} }$ is deterministic.
In contrast, the operational cost $c_{s, \text{operational}}$ for shared service is an expectation conditional on whether the system will match this ride with others in the future and how the cost is split, and is computed as follows:
\begin{equation}\label{equ: expec_cost}
     c_{s, \text{operational} } = \left(1 - \sum\limits_{r' \in \mathcal{R}_{m}} \alpha_{r'}\right)  c_0 + \sum\limits_{r' \in \mathcal{R}_{m}} \alpha_{r'} \left(c_{r', r} - p_s^{r'} \right),
\end{equation}
where $c_0$ is the operational cost if the request is not matched with any requests in the future; $\mathcal{R}_m$ is the set of all possible future requests for mode $m$ that might be matched to the request $r$; $\alpha_{r'}$ is the probability that request $r'$ is matched to $r$ conditioned on the customer submitting $r$ choosing the shared option; $c_{r', r}$ is the additional cost of adding request $r'$ to the vehicle matched to $r$, which is the difference between the cost of serving both requests and the cost of serving only the first request; and $p_s^{r'} $ is the price of request $r'$.
Although the distribution for each $r' \in \mathcal{R}_m$ cannot be computed in exact form, expected operational costs can be estimated empirically. 

The retrospective cost $c_{ \text{retrospective} } $ penalizes a suboptimal assignment (i.e., one in which the vehicle could have been matched to a better request in the future). 
When a vehicle $v$ is assigned to serve a request $r$ (or multiple requests if $v \in \mathcal{V}_s$), it is blocked from future operations within a given duration.
The retrospective cost considers the spatio-temporal distribution of requests that occur during this time period, because the supply levels at both origin and destination areas are affected by a suboptimal assignment. 
The actual retrospective cost is assumed to follow a function $g_{\pi_p, \pi_{a}}(\mathcal{R}_m)$.  
A surrogate function $\hat{g}(\cdot)$ is used as a proxy for the retrospective cost because the exact spatio-temporal demand distributions are not available. 
Any advanced estimators for the continuation value of future requests $\mathcal{R}_m$ can be easily integrated into this pricing framework.
For example, TNCs have used recurrent neural networks to predict the profit and estimated time-of-arrival in real-time \parencite{li2019efficient,shah2020neural}.

This work proposes a straightforward estimator $\hat{g}(\cdot)$ that requires minimal access to contextual demand and supply data.
We divide a fixed horizon into $M$ time intervals and the road network into $N$ clusters. 
For each time interval and cluster, we compute the average profit made per unit time and per vehicle in the cluster based on the steady-state pricing model \parencite{castillo2017surge} and extend it to the ridepooling case.  
The MoD system follows supply flow balance equations: 
\begin{align}
	& \quad L = L_e + L_s,  \label{eq6} \\
	& \quad Y = Y_e + Y_s,   \nonumber 
\end{align}
where $L$ is the total number of vehicles in the cluster at the current time interval, and $Y_e$ and $Y_s$ are the steady-state throughput of exclusive vehicles and sharing vehicles, respectively. 

The total number of vehicles per type can be calculated as:
\begin{align*}
     \begin{cases}
		L_e =  O_e + \eta_e Y_e + T_e  Y_e \\
		L_s =  O_s +  \frac{\eta_s   }{\zeta_s }    Y_s +  \frac{  T_s  }{\zeta_s}  Y_s
	\end{cases},  \nonumber 
\end{align*}
where $O_e$ and $O_s$ are the number of available vehicles of each group, and $\eta_e$ and $\eta_s$ are the average waiting times for the customers choosing exclusive and shared services. The expected waiting time is assumed to be equal in each group because the assignment policy $\pi_{a}$ is fixed.  
$T_e$ and $T_s$ represent the average trip duration of each group at steady-state.
According to Little's law, $\eta Y$ represents the number of vehicles in pick-up trips and $T \cdot Y$ represents the number of en-route vehicles. 
$\zeta_s$ is the average utilization of sharing vehicles in ridepooling over time. We explain how to calculate the average utilization in Appendix~\ref{sec:app2}.

The waiting times $\eta_e$ and $\eta_s$ are determined by the density of available vehicles near the origin location \parencite{zha2018geometric}. 
Without loss of generality, we assume $\eta_e = F_e(O_e)$ and $\eta_s = F_s(O_s)$, and obtain the approximate trip throughput as: 
\begin{align}
    &Y_e(\eta_e, O_e) \triangleq \frac{L_e - O_e}{F_e(O_e) + T_e}, \nonumber  \\
    & Y_s(\eta_s, O_s) \triangleq \frac{ \zeta_s ( L_s - O_s)}{F_s(O_s) + T_s}.
\end{align}


For a request $r$ traveling from the origin region $o$ to the destination region $d$, we can compute the average profit obtained in each region by combining the throughput $Y_e, Y_s$ and listed prices $p_e, p_s$. For each region $k$, $\bar{\epsilon}^k_e$ and  $\bar{\epsilon}^k_s$ are the respective average profits. 
Given the trip duration $t_r$ for an exclusive service vehicle, the retrospective costs $c^{\pi_p}_{m, \text{retrospective}}$ for mode $m\in \{e,s\}$ are as follows:
\begin{align} \label{equ: potential}
    c_{m, \text{retrospective}} & = \hat{g}(t_r, t_r') =  \bar{\epsilon}^{o}_m  t_r + ( \bar{\epsilon}^{o}_m - \bar{\epsilon}^{d}_m )  t_r^\prime, 
\end{align}
where  $\bar{\epsilon^{o}_m } \cdot t_r$ is the type $m$ vehicle's average profit collected from staying in the origin $o$, and 
$(\bar{\epsilon}_m^{o} - \bar{\epsilon}_m^{d}) \cdot t_r^\prime$ is the average profit difference that the vehicle makes between the two clusters in time $t_r^\prime$. 
The computation of retrospective cost connects the supply and demand information across the MoD network. 
A similar approach has been used in modeling taxi and ride-hailing vehicles' movement \parencite{yang2002demand, bimpikis2019spatial}; the derivation of these approximate forms are based on a steady-state analysis of the MoD system dynamics.

\subsubsection{Multi-product pricing (MPP) problem for mixed fleets.}

The expected profit of the MPP problem under the SPD framework is:
\begin{equation}\label{equ: obj}
    \mathbb{E}_{\pi_p, \pi_{a} }  [\Phi_r]  = \frac{ e^{  \beta_{p} p_{e} +  U_e^{\pi_{a}} }  (p_{e} - c_e) +  e^{  \beta_{p}  p_{s} +  U_s^{\pi_{a}} }  (p_{s} - c_s)   }{  e^{  \beta_{p} p_{e} +  U_e^{\pi_{a}} } + e^{  \beta_{p}  p_{s} +  U_s^{\pi_{a}} }  + e^{  \beta_{p}  p_{o} +  U_o^{\pi_{a}} }   }.
\end{equation}

In general, the objective function \eqref{equ: obj}  is not convex in prices \parencite{hanson1996optimizing}. 
A common convexification technique is to project the pricing to market shares and transform the objective function \parencite{li2011pricing} as follows. 
For the special case where each vehicle serves at most two rides per trip, we show a closed-form expression of unique optimal prices.
 
\begin{proposition}\label{prop: seq}
    The objective function has a unique critical point given by
    \begin{equation}
        p_e^* = \frac{1}{\beta_p}   \left\{ \log \Bigg[ \dfrac{ e^{U_0}  \cdot W\left( \frac{ (1 + e^{U^{\pi_{a} }_{s} - U^{\pi_{a} }_{e} + \beta_p c_s - \beta_p  c_e }) \cdot e^{U^{\pi_{a} }_{e} + \beta_p c_e - 1} }{  e^{U_0} }  \right)   }{1 + e^{U^{\pi_{a} }_{s}   - U^{\pi_{a} }_{e} + \beta_p c_s -  \beta_p c_e  }} \Bigg] -  U^{\pi_{a} }_{e}   \right\},
    \end{equation}
    where $W(\cdot)$ is the Lambert W function, and 
    \begin{equation*}
        p_{e}^* - p_{s}^* = c_e  - c_s.
    \end{equation*}
\end{proposition}

\begin{remark}
	 The difference between the price for exclusive service and the price for shared service at the critical point is the expected cost savings from choosing the shared service.
\end{remark}

Since the Lambert W function is positive and increasing when the input is nonnegative, the objective function~\eqref{equ: obj} has a unique solution. 
Hence, the optimal price for two services corresponding to an arbitrary origin-destination pair can be computed by checking all the boundary points and the critical point giving the maximum profit. 
The proof of Proposition~\ref{prop: seq} can be found in Appendix~\ref{proof:prop:seq}. 

In summary, the MoD platform can present the prices computed by the SPD framework to each arriving customer so that she will choose the mode that maximizes her utility (or opts out for an outside alternative). 
The SPD framework can be implemented in real time as each service's optimal price has a closed-form solution.


\subsection{Pricing in BPD Framework}

This section presents how to set prices under the BPD framework. 
Since BPD holds upcoming trip requests in batches and optimizes each batch's operations, it is more advantageous for a platform seeking to improve the matching quality, reduce the overall operational costs, and optimize routes for multiple pickups and dropoffs \parencite{ke2020ride}.
This section focuses on a BPD framework with a vehicular capacity of two, which is later extended to higher capacity cases.\footnote{Most TNCs currently operate a low-capacity services due to the capacity of their vehicle fleet and other factors \parencite{ke2021data}.} 
The main challenge is to jointly determine the trip-vehicle assignment and prices for each batch.    

\subsubsection{Vehicle dispatch with overbooking.}


The construction of the shareability graph in \citet{santi2014quantifying}, which describes the matchable relationship between available vehicles and pending requests, needs to be expanded to consider mixed service types. 
However, 
extending the formulation in \citet{alonso2017demand} to consider endogenous future demand is non-trivial.  
The main research question is how to incorporate the value of reserving supply or demand in the exact optimal assignment to maximize the cumulative profit in batches.
At a high level, we first construct a Request-Vehicle (RV) graph and convert it to a Request-Trip-Vehicle (RTV) graph as in \citet{alonso2017demand}. 
Finally, we can create a so-called Exclusive-Sharing-Vehicle (ESV) graph that accounts for both exclusive and shared services. 
These graphs describe the feasible assignment between vehicles and pending trip requests. 
Subsequently, we solve a global integer optimization to find the optimal prices and trip assignment for maximizing expected profit.



\begin{figure}[!htb]
	\hspace{-0.3in}
	\subfloat[RV graph \label{fig2-1} ]{ \includegraphics[width = 0.33\textwidth]{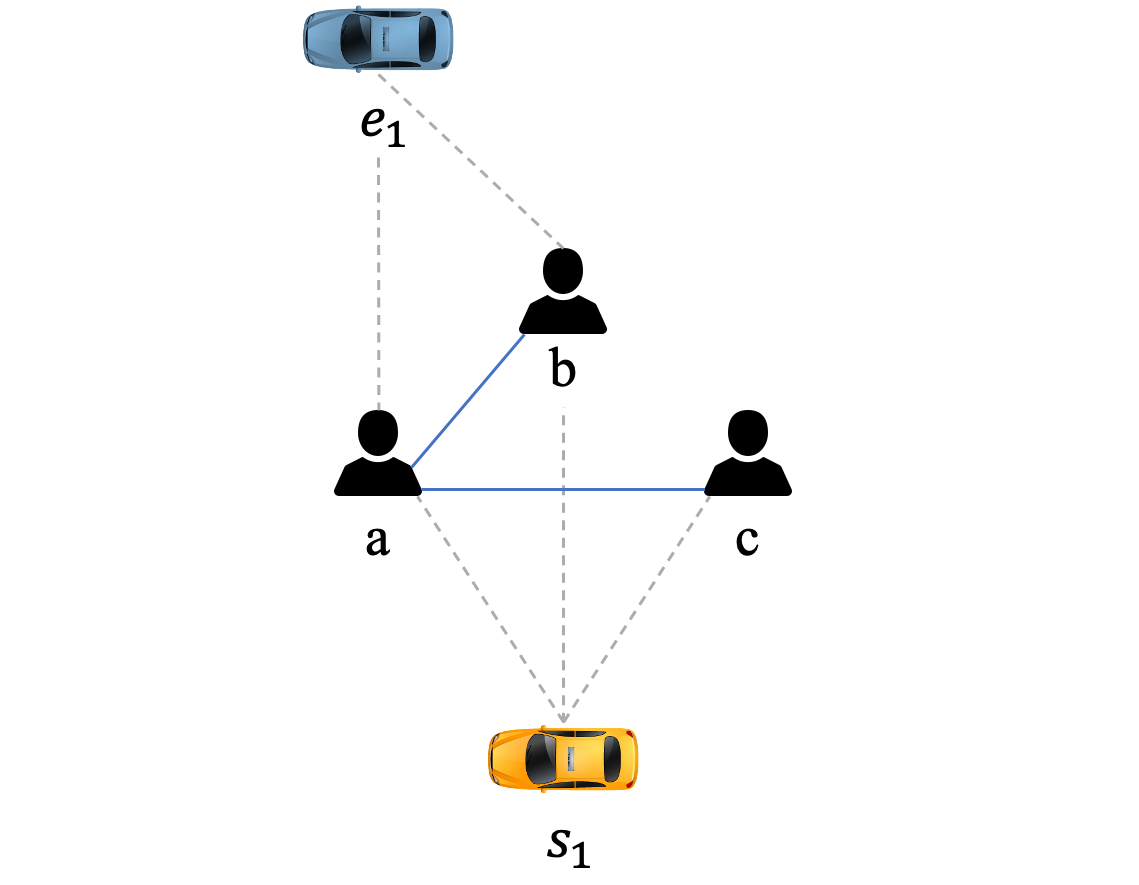} }
	\subfloat[RTV graph  \label{fig2-2} ]{ \includegraphics[width = 0.33\textwidth]{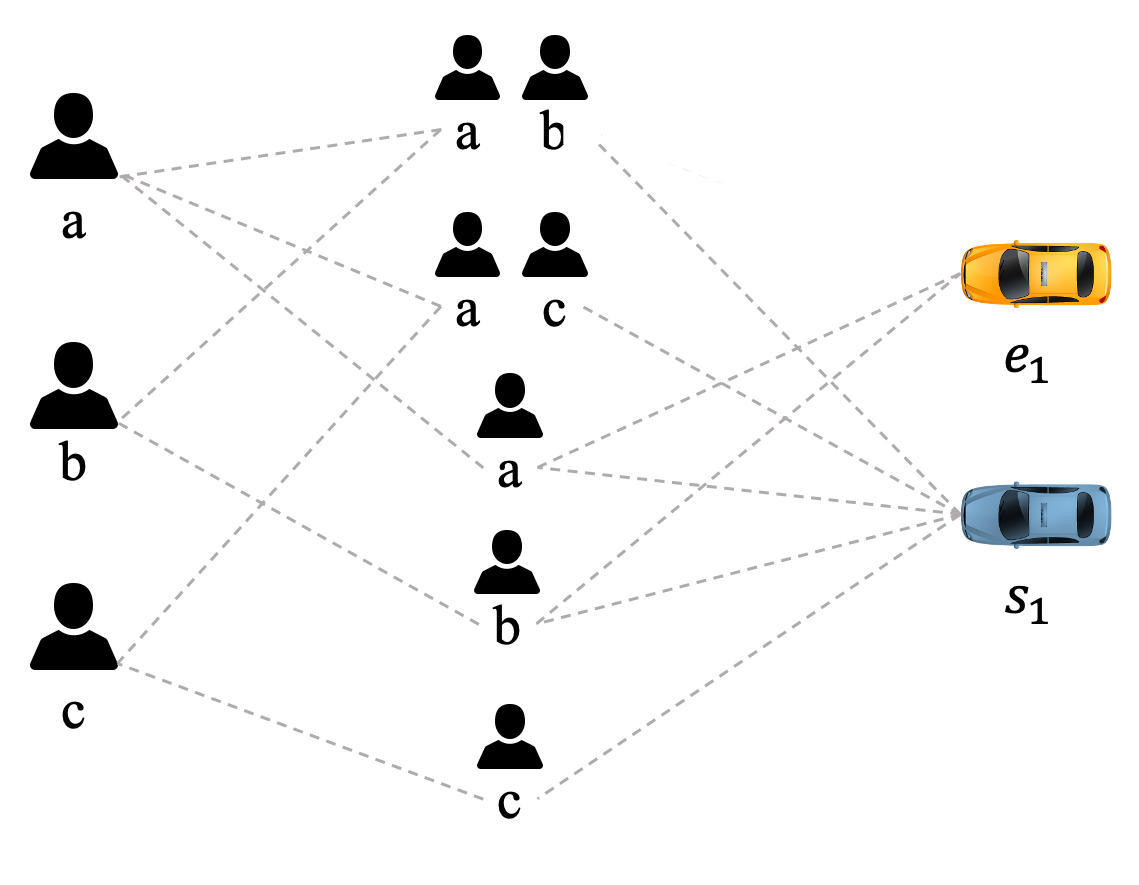} }
	\subfloat[ESV graph  \label{fig2-3} ]{ \includegraphics[width = 0.33\textwidth]{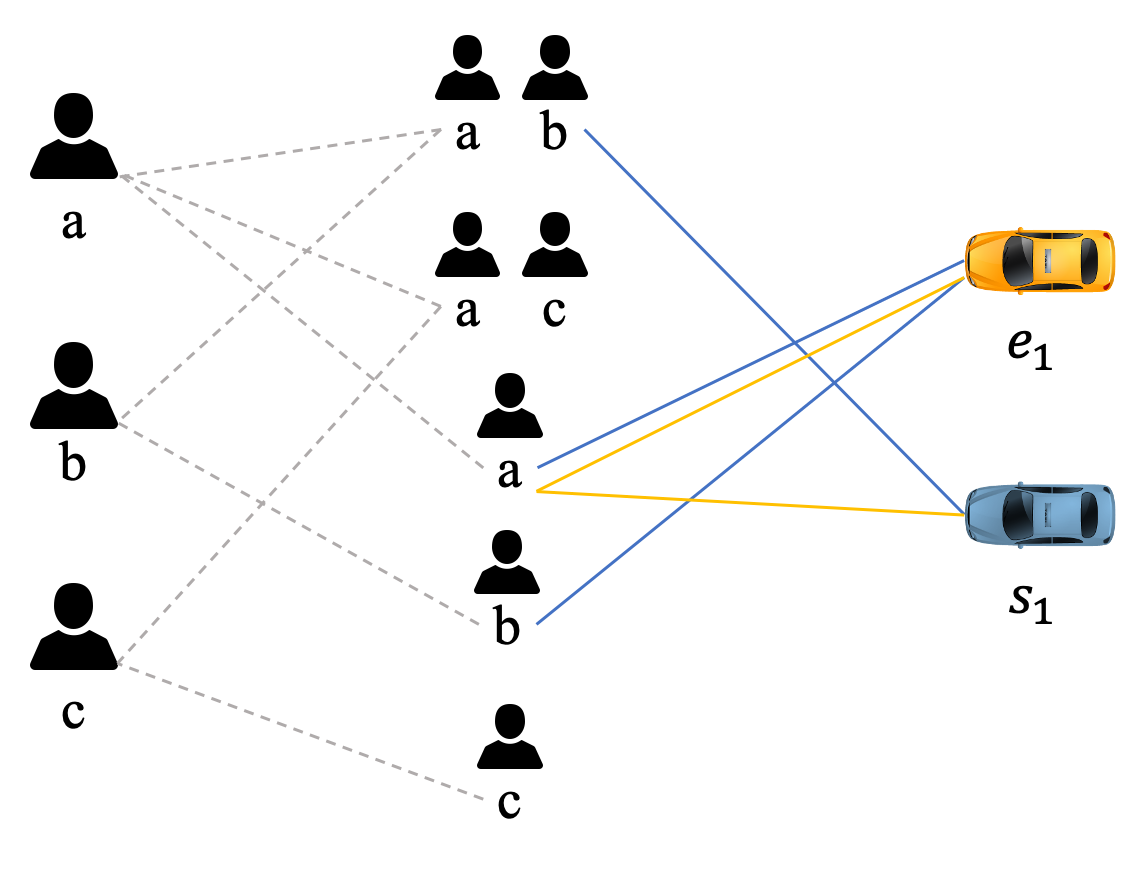} }
	\caption{Constructing a shareability graph in  batched pricing; $e_1$ is an exclusive vehicle and $s_1$ is a sharing vehicle; We show two sample ESV matchings containing one and two requests, respectively. } \label{fig2}
\end{figure}

More concretely, given a batch of supply (available vehicles $\mathcal{V}$) and demand (trip requests $\mathcal{R}$), the procedure is as follows: 
\begin{enumerate}
	\item Construct a RV graph (Figure \ref{fig2-1}). 
	The RV graph gives all possible pairwise matchings between vehicles and requests. 
	In this graph, two requests $r_i$ and $r_j$ or a request $r$ and a vehicle $v$ are connected if the conditions discussed in Section \ref{sec212} are satisfied.  
	
	\item Construct a RTV graph (Figure \ref{fig2-2}) using the matchings in the RV graph. 
	A trip $T$ is a clique of requests that can be served by a vehicle without violating any feasibility constraints.  
	
	\item Compute an ESV graph (Figure \ref{fig2-3}) graph using both the RV graph and the RTV graph. 
	Each ESV matching contains either one request or two requests:
	\begin{itemize}
		\item  If the ESV matching contains one request, the clique contains a combination of the following vehicles based on the RV graph: 
		\begin{enumerate}
		\item An exclusive service vehicle $v_e$ available for this request.
		\item A shared service vehicle $v_s$ available for this request. 
		\end{enumerate}
		
		\item 	If the ESV matching contains two requests, the clique contains a combination of the following vehicles based on both the RV graph and the RTV graph:
		\begin{enumerate}
			\item A shared service vehicle $v_s$ that is able to serve these two requests.
			\item An exclusive service vehicle $v_e$ available for the first request.
			\item An exclusive service vehicle $v_e'$ available for the second request. 
		\end{enumerate}
	\end{itemize}
\end{enumerate}

\noindent 

Compared to the standard TVA formulation in \parencite{alonso2017demand}, we need to relax the packing constraints in the MIP to include the customers' mode choice behavior, as follows: 
\begin{subequations}\label{eq:BDP}
	\begin{align}
			\text{maximize: }  & \displaystyle\sum\limits_{i \in \mathcal{M}}   u_{i} y_{i} - \sum\limits_{j \in \mathcal{V}} w_j  \tag{\ref{eq:BDP} }  \\
			\text{subject to: }& \displaystyle \sum\limits_{i \in \mathcal{I}_{R = k}^{\mathcal{M}}}   y_i \leq 1,   & \forall k \in \mathcal{R}  \label{eq:BDPa} \\ 
			&w_{j} \geq (c_p + \bar{ \epsilon}_j) \cdot (\sum\limits_{i \in \mathcal{I}_{j }^{\mathcal{M}}} y_i  \gamma_{i, j} - 1), &  \forall j\in \mathcal{V}  \label{eq:BDPb}  \\ 
			&w_{j} \geq 0, 		&  \forall j\in \mathcal{V}  \label{eq:BDPc}   \\ 
			& y_{i} \in \{0,1\}, 	&  \forall i\in \mathcal{M}   \label{eq:BDPd} 
	\end{align}
\end{subequations}
where $u_{i}$ is the expected profit of the ESV matching $i$, and $\mathcal{M}$ is the set of ESV matchings. 
The binary decision variables $y_i$ represents whether the system choose an ESV matching $i \in \mathcal{M}$.  
In the packing constraints Eq.~\eqref{eq:BDPa}, we split the potential ESV matchings into two sets: (a) a set of matchings that is connected to vehicle $j$ as $\mathcal{I}_{j}^{\mathcal{M}}$, 
and (b) a set of matchings that contains request $k$ as $\mathcal{I}_{k }^\mathcal{M}$.
This packing constraint ensures that each request is assigned to at most one vehicle.

Since the mismatch of vehicle and requested service type is possible in the BPD framework, we adopt an overbooking strategy to enhance the overall performance in Eq.~\eqref{eq:BDPb}. 
The platform with mixed fleets allows each vehicle to be assigned to multiple ESV matchings.  
$\gamma_{i, j}$ is the probability that, as the ESV matching $i$ containing the vehicle $j$ is selected, vehicle $j$ is used at the end. 
If vehicle $j$ is designated for the exclusive service, $\gamma_{i, j}$ is the predicted probability of the corresponding customer accepting the trip.
If vehicle $j$ is designated for the shared service, $\gamma_{i, j}$ is the probability that at least one of the assigned customers accepts the trip.
If more than one customer has accepted the assignment, then at least one request is either reassigned or rejected when there is no available vehicle nearby. 
We impose a fixed penalty $c_p$ for each lost demand.  
Let $\bar{\epsilon}_j$ be the mean expected profit earned by vehicle $j$ among all feasible matchings.  
Variables $x_j$ are the overbooking penalty for vehicle $j \in \mathcal{V}$. 
However, the system needs to evaluate an exponential number of scenarios to calculate the exact number of lost demands.  
Therefore, we use $\sum_{ i \in \mathcal{I}_{j }^{\mathcal{M}} } y_i \cdot \gamma_{i, j} - 1$ as an approximation and simplify this computation.
The BPD prices are computed using the approach in Section \ref{sec: batchpricing} and the maximum waiting time and delay for each request is updated for each batch. 


\subsubsection{Price mechanism for ESV matchings.}\label{sec: batchpricing}
In this section, we show how to find the optimal prices for each trip request in ESV matchings. 
If the matching contains only one request, the optimization procedure is the same as shown in the SPD framework. 
Therefore, we focus on the case where the ESV matching consists of two requests. 

Suppose that $r_1$ and $r_2$ are the two requests in an ESV matching. The expected profit function of MPP is given by: 
\begin{equation}\label{equ: batchprofit}
    \begin{split}
    	\mathbb{E}_{\pi_p} [ \Phi_{r_1, r_2} ] &= P_{1, s}  (p_{1, s} - c_{1, s}) + P_{2, s} (p_{2, s} - c_{2, s}) + P_{1, e}  (p_{1, e} - c_{1, e})  \\ 
    	&+ P_{2, e}  (p_{2, e} - c_{2, e}) + P_{1, s} P_{2, s}  (c_{1, s} + c_{2, s} - c_{s, s}),
\end{split}
\end{equation}
where $p_{1, s}$ and $p_{1, e}$ are the shared service price and exclusive service price for $r_1$, respectively. 
$p_{2, s}$ and $p_{2, e}$ are prices for $r_2$.  
$c_{1, e}$ and $c_{2, e}$ are the cost incurred in serving $r_1$ and $r_2$ exclusively.  
$c_{1, s}$ and $c_{2, s}$ are the expected cost of serving $r_1$ and $r_2$  if they do not form a sharing trip. 
$c_{s, s}$ is the cost of serving $r_1$ and $r_2$ in a single sharing trip. 
The cost is an expectation that depends on whether these requests are shared with other requests and whom to be shared with. 

We have the following property of the objective function in regard with the probabilities $Pr_{1, e}$, $Pr_{1, s}$, $Pr_{2, e}$, and $Pr_{2, s}$:
\begin{proposition}\label{prop: bat}
    The objective function Eq.~\eqref{equ: batchprofit} is jointly concave in the predicted probabilities if 
    \begin{equation}
        \min\{c_{1, e}, c_{2, e}\} \leq -\frac{1}{\beta_p}.
    \end{equation}
\end{proposition}

The proof is provided in Appendix \ref{proof: prop: bat}. 
If this condition is not satisfied, we use a brute-force method to find the optimal solution for the nonconvex profit function. 
Specifically, we can enumerate $P_{1, s}$ in the range $[0, 1]$ with a predefined step length. 
Given the value of $P_{1, s}$, we can prove that the problem to optimize $P_{1, e}$, $Pr_{2, s}$ and $P_{2, e}$ is jointly concave in the variables using similar methods. 
After enumerating all $P_{1, s}$, we can return the prices with the highest expected profit.

\section{Numerical Experiments}\label{c4sec: experiments}
\subsection{Overview of NYC case study}

The data used to estimate the model were obtained from a stated-preference study in NYC. 
We use the publicly available Taxi and Limousine Commission (TLC) trip record data in Manhattan, NYC as a proxy for the real O-D demands, which is an established procedure in literature \parencite{santi2014quantifying, alonso2017demand}. 
In this market, an MoD platform operates mixed fleets and competes for customers who use the existing taxicab service.  
This numerical case study uses taxicab trip data as the requests $\mathcal{R}$ for both MoD and outside services.

We run the simulation using multiple consecutive weekdays' demand data (April 1, 2013 to April 9, 2013) in Manhattan, NYC.
To serve the travel demand in the area of study, the platform operates a fleet of 2000 vehicles with capacity one as $V_e$ and 1000 vehicles with capacity two as $V_s$. 
The road network in Manhattan, NYC consists of 4092 nodes and 9453 edges. 
The edge travel time is estimated by the daily mean travel time using the method in \citet{santi2014quantifying}. 
The pickup time in the taxicab trip data is considered as the starting time of each trip request. 
The reader interested in more details on the MoD simulator can refer to the previous work \parencite{alonso2017demand}.

All aforementioned methods are implemented using Python $3.5$, and all experiments are conducted on an Intel Core i7 computer (3.4 GHz, 16 GB RAM). 
In this section, we present a series of numerical experiments to show: 
\begin{enumerate}
	\item The profit improvement by considering retrospective costs.
	\item The performance comparison between SPD and BPD.
	\item The profit increase of our framework compared with  static pricing methods (discussed in Section~\ref{sec: calibration}). 
\end{enumerate}

\subsection{Data Description and Parameter Estimation}
We estimate the parameters in the implementation of SPD and BPD frameworks in three main steps. 
First, most parameters in the pricing policies can be estimated from historical data. 
Second, we check main modeling assumptions a) the convexity of the objective and b) the convergence of mean costs to the derived analytical forms. 
Finally, we calibrate the mode choice model. 

\subsubsection{Estimation of parameters in pricing policies.}

The average trip duration, trip costs, and the customers' arrival rate at each time interval are computed from historical data.
Currently, TNCs mainly hire freelancer drivers who receive a fixed proportion of fares as their commission fees. 
However, since our operational setting assumes a centrally controlled vehicle fleet (e.g., automated vehicles) that follows dispatching and re-positioning guidance from the platform, driver compensation are not considered throughout this experiment.

Next, we calibrate the functions $F_e(\cdot)$ and $F_s(\cdot)$.
Since the system's supply and demand are equal at the equilibrium, we can obtain the prices for both services given the number of open vehicles and compute the expected total profit.
In summary, given the total number of vehicles for the exclusive service and the shared service $L_e$ and $L_s$, we can enumerate $O_e$ and $O_s$ and compute the corresponding prices $p_e$ and $p_s$, respectively. 

Finally, we calculated the prices offered to customers by the developed heuristic because it is difficult to determine $t_r^\prime$ at the request level. 
The prices are calculated as maximizers to the average profit per period for each vehicle providing services $\bar{\epsilon}_e$ and $\bar{\epsilon}_s$.
The period is set it to be 20 minutes throughout the experiment.


\subsubsection{Test of modeling assumptions.  }
Two critical assumptions are tested in this numerical experiment. First, whether the objective function is jointly concave; Second, if cost functions converge to their analytical forms. 

\noindent \textbf{Test of concavity in Proposition \ref{prop: seq}.} \quad  
According to our estimated mode choice model, $-1/\beta_p \approx 13.5$ in the numerical experiment. 
In other words,  if the average operational cost in each ESV matching is smaller than \$13.5, the problem is jointly concave in the predicted probabilities.

The area of interest in our experiment, Manhattan borough, has an area of 22.7 square miles (13.4 miles long and 2.3 miles wide at its widest). 
If the MoD carriers are automated vehicles, the driving cost is about \$0.1458 per mile including the cost of fuel, maintenance and tiers \parencite{liu2018framework}. 
Therefore, if the shorter trip in each ESV matching, $r_1$ and $r_2$, is less than 92.6 miles, the problem is jointly concave in the predicted probabilities and we can compute the unique optimal prices for each request. 

\noindent \textbf{The convergence of expected costs.} \quad  
As mentioned in Eq.~\eqref{equ: expec_cost}, if a request is the first being assigned to a sharing trip, the operational cost is an expectation that depends on whether and who this request will be matched with. 
Since the expected cost does not have an explicit form, we run simulations to obtain an empirical estimator. 
Specifically, we test the convergence by the following procedure: 
\begin{enumerate}
	\item As the demand is sparsely distributed in the road network, we use the K-means clustering algorithm to cluster the nodes according to their geo-coordinates spatially using the method in \citet{liu2020proactive}. 
	Assuming that each customer has a walking range of $\alpha$ miles, we choose the number of clusters $K$ such that these clusters cover the entire area approximately. 
	\item  The simulation runs for a long period (e.g., a week) and aggregates these trips into O-D pairs between these clusters.
	\item  Operational costs for each trip are recorded assuming that travel demand is exogenous and unaffected by the implemented policies.  
\end{enumerate}

The expected cost for each O-D cluster is initialized to be 0 at the beginning. 
We compute the average realized cost for each pair of clusters at the end of simulation, and use it as the expected operational cost for each O-D pair that belongs to these clusters.
After running each day's simulation, we update the average cost for each O-D cluster based on completed results, and use it as the starting value for the next day.
Although the expected cost and the mode choice decisions in the simulation are coupled, the expected cost between each cluster can converge to a relatively stable throughout the simulation.

\begin{figure}[!htb]
\centering
\includegraphics[width=0.65\textwidth]{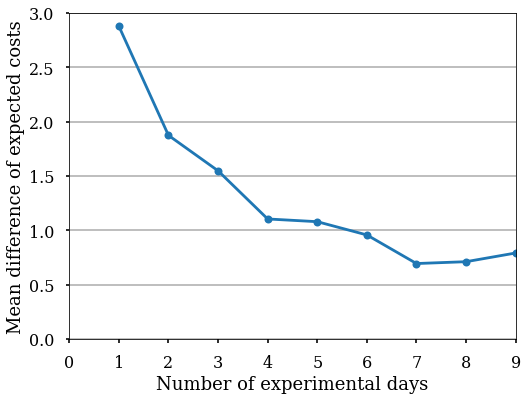}
\caption{\label{fig: MAD_cost} The mean absolute difference of the expected cost among O-D clusters.}
\end{figure}

Figure~\ref{fig: MAD_cost} shows the mean absolute difference of the expected cost among all O-D clusters between each day and the previous day. 
The curve has a decreasing trend and fluctuates around a relatively low value, which indicates the numerical convergence of the expected cost. \par 

\subsubsection{Mode choice model calibration.} \label{sec: calibration}
We use a \emph{static} price adopted by UberX service in the past \parencite{Uber} as a benchmark throughout this experiment. 
The price is computed by:
\begin{equation}\label{equ: trip_cost}
    p_r = \max (f_{min}, f_{base} + f_t  t_r + f_d  d_r)
\end{equation}
where $f_{min}$ and $f_{base}$ are the minimum fare and base fare for the ride-hailing service, $t_r$ and $d_r$ are the travel time and travel distance for $r$, $f_t$ is the time rate per second for the ride-hailing service, and $f_d$ is the distance rate per mile for the ride-hailing service. 
We use the same coefficients from a previous NYC Uber empirical study \parencite{Uberprice}. 

\begin{figure}[!h]
\centering
\includegraphics[width=0.65\textwidth]{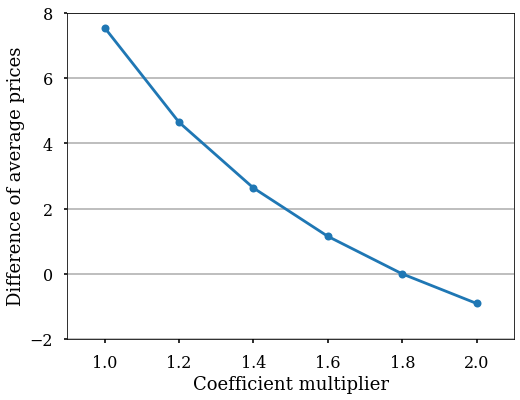}
\caption{\label{fig: utility_scale} The price comparison between the static price and the price of our framework under different coefficient multipliers.}
\end{figure}

The choice model that reflects the customers' preference between MoD and conventional taxi service is calibrated as in the prior work \parencite{liu2018framework}.
Customers choosing between MoD services and public transit mainly consider the QoS, such as total travel time and variations of waiting time. 
In comparison, customers are more price-sensitive as MoD and taxi service are similar in this dense urban network. 
Practitioners may conduct extra surveys to estimate the mode choice model's coefficients more accurately. 

We circumvent the issue with parameter errors by incorporating a multiplier greater than 1 to scale the price coefficients in the MNL mode choice model. 
We simulate the pricing-and-assignment process using an arbitrary day's data (April 10, 2013) using different multipliers  $\{1.2, 1.4, 1.6, 1.8, 2.0 \}$. 
The goal is to find a multiplier such that the average price determined by our framework is similar to the average static price computed by Eq.~\eqref{equ: trip_cost}.
Figure \ref{fig: utility_scale} shows the average price between applying the proposed SPD/BPD framework and that of the static price benchmark. 
Since the multiplier of 1.8 gives the same average price as the average static price, we fix it in the remainder of the experiments and rerun the above experiments to estimate O-D clusters' expected operational costs.

\subsection{The Effects of Retrospective Cost Function}
Retrospective costs represent the expected sunk profit of assigning an available vehicle to future requests. 
In this section, we use the simulation of demand data on April 16, 2013, to illustrate the effect of including the retrospective cost in Eq.~\eqref{equ: potential}. 
The overall profit decreases after employing retrospective costs directly.
Possible reasons include: 
\begin{enumerate}
	\item  Eq.~\eqref{equ: potential} is based on a steady-state analysis  \parencite{castillo2017surge,yan2020dynamic}. 
	The actual demand and supply processes are nonstationary, and hence $\hat{g}(\cdot)$ is not a precise estimator for $g(\cdot)$.
	\item  The retrospective cost represents the predicted profit a vehicle could make during the travel time. 
	This estimation may not be accurate in a clustered network as the travel time may vary significantly from individual requests. 
\end{enumerate}

\begin{figure}[htb]
\centering
\includegraphics[width=0.65\textwidth]{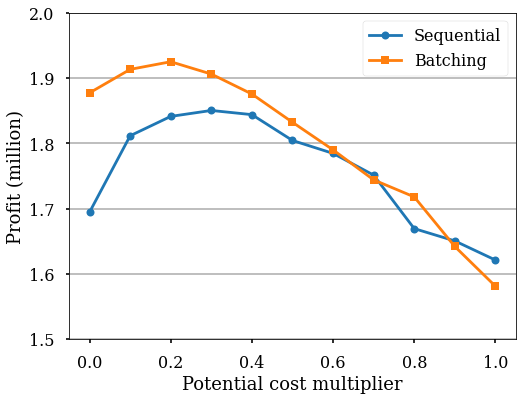}
\caption{\label{fig: opportunity} The profit of both the batching pricing framework and the sequential pricing framework under different retrospective cost multiplier. }
\end{figure}

As mentioned in Section \ref{sec: batchpricing}, finding the exact form of $\hat{g}(\cdot)$ is not the focus of this work. 
We conduct a sensitivity analysis using a multiplier to scale the retrospective cost and demonstrate the effect of including retrospective costs.
The multiplier ranges from 0.0 to 1.0 with a step of 0.1 in the following experiment. 
We run the experiments separately for each instance and use multipliers that give maximum total profits. 
Note that optimal multiplier may differ for the BPD framework and the SPD framework.

Figure~\ref{fig: opportunity} shows profits of both frameworks under different retrospective cost multipliers.
There exists an optimal multiplier larger than $0$ that maximizes the expected profit (0.2 and 0.3 for the batching pricing framework and the sequential pricing framework, respectively). 
The total profits increase by 2.7\% and 9.2\% for the BPD framework and the SPD framework, respectively.
The introduction of retrospective costs has less impact on the BPD framework because the setup of batched requests has already considered the partial effect of delayed reward. 

\subsection{Performance Comparison}
In this section, we compare the performance of our BPD and SPD frameworks and the results are presented in Table~\ref{c4: model_compare}. 
The simulation uses the demand data on April 17, 2013, in Manhattan, NYC. 
As this work intends to improve the performance of high-capacity MoD dispatch algorithms in \citet{alonso2017demand}, we use two benchmark models that employ a static pricing mechanism (i.e., fixed rules of splitting trip fares in ridepooling) in the TVA formulation. 

In the static pricing method, the fractions of customers choosing shared services are calculated by the probability of the customer sharing with other requests in the future.
We assume that the shared service's fares are split based on the corresponding exclusive service prices. 
Let $\kappa$ be the discounter of the shared service and $\theta$ be the probability of the request $r$ sharing with other requests in the future, which is computed empirically using simulation on historical data for each O-D cluster. 
Then, given the exclusive service price $p_e$, we compute the shared service price $p_s$ as $  p_s = (1 - 0.3 \cdot \theta + 0.2) \cdot p_e$ \parencite{Uberprice}.
This simple price structure can be easily calibrated with access to real-world ride-hailing trip data.

Table~\ref{c4: model_compare} lists the profit, the market share, the mean price, and the mean waiting time of customers. 
The overall profits of the BPD framework and SPD framework increase significantly and decrease the mean waiting time compared to the corresponding static pricing method (23.6\% increase and 39.1\% increase). 
However, the market shares decrease by 8.0\% with the BPD framework and 6.6\% with the SPD framework, respectively. 
One explanation is that our pricing frameworks set the prices higher than the status-quo static pricing mechanism. 
Although the market share decreases, the profit per trip grows significantly, and total profits increase with either framework.

\begin{table}[!htb]
	\caption{Model comparison}
	\label{c4: model_compare}
	\resizebox{\textwidth}{!}{%
	\begin{tabular}{lcccc}
			\toprule
			Model & Profit (million) & Market share (\%) & Mean price & Mean waiting time (s) \\
			\midrule
			BPD & 1.99 & 34.54 & 14.37 & 135.60 \\
			Batched-Static & 1.61 & 42.49 & 9.73 & 144.52 \\
			SPD & 1.92 & 32.58 & 14.72 & 117.88 \\
			Sequential-Static & 1.38 & 39.15 & 9.44 & 119.56 \\ 
			\bottomrule	
	\end{tabular} 
	} 
	
    \smallskip 
	{\small Batched-Static: Matching the requests in batches and set prices using the static price in Eq.~\eqref{equ: trip_cost}. \\
	\noindent Sequential-Static: Matching the requests sequentially and set prices using the static price in Eq.~\eqref{equ: trip_cost}.
	}
\end{table}

Compared to the SPD framework, the BPD frameworks can increase profits due to more efficient matching and pricing in batches. 
Specifically, the BPD framework increases the profit by about 3.6\% compared to the SPD framework, and the profit of the \textit{Batched-Static} framework is 16.7\% higher than the \textit{Sequential-Static} framework. 
The profit increase of BPD is at the expense of an increase in the mean waiting time because the batched supply and demand uses a 30-second matching window. 
In practice, the MoD platform should find the optimal trade-off between the overall service level and profits.

\section{Conclusion} \label{sec:conclusion}

This study proposes a computationally efficient approach for solving the JVDPP on an MoD platform. 
The platform providing exclusive and shared services must continuously balance each option's QoS and prices to accommodate various demands and maximize the total profit.
An important step towards operating such a platform is to design a scalable dispatching and pricing method with modest performance guarantees. 
Leveraging analytical MPP and spatial pricing results, we propose a sequential framework and a batched framework that simultaneously compute prices and dispatch ridepooling trips at the request level.

We show a closed-form solution to the optimal prices in the SPD framework and prove the convexity of the BPD's objective with regard to mode choice probabilities  under mild technical assumptions on the cost structures. 
Based on these results, we propose a scalable MIP formulation with overbooking to overcome the myopicity in the BPD framework. The notion of retrospective costs, a spatiotemporal proxy to the suboptimal assignment, can increase the overall profit by about 3\% to 9\% in numerical experiments. 
The SPD/BPD framework yields significant profit increases (24\% to 39\%) compared to the benchmark static pricing method. 

There are multiple avenues for future research. 
First, the tractability of this model is based on MPP results for up to two requests. 
It is worth investigating how to improve the brute-force search procedure for high-capacity MoD fleets with mixed fleets. 
Second, considering more realistic cost structures with advanced data-driven approaches than the steady-state analysis and reevaluating SPD/BPD frameworks' effectiveness is a promising direction. 

\section*{Acknowledgement}
This research is supported by NSF DMS-1839346 and SCC-1952011.

\clearpage

\ECSwitch


\ECHead{Proofs and supplementary materials}

\begin{appendices}

\section{Summary of notation} \label{sec:app1}
We summarize the notation used throughout this work in the following table:
	\begin{center}
	\begin{longtable}{c|l}
		\caption{Summary of notation}   \label{table:app1} \\
		\toprule
		\multicolumn{1}{c|}{Notation} & \multicolumn{1}{c}{Definition}               \\ \midrule
		\multicolumn{2}{l}{Terminology}                                                         \\ \midrule
		MoD & Mobility-on-Demand \\ \hline 
		TNC & Transportation network company \\ \hline 
		JVDPP & Joint vehicle dispatching and pricing problem \\ \hline
		SPD & Sequential pricing and dispatch \\ \hline 
		BPD & Batched pricing and dispatch \\ \hline 
		MIP & Mixed integer programming \\ \hline 
		MNL & Multinomial logit model \\ \hline 
		O-D pair & Origin-destination pair \\  \hline
		QoS & Quality of service \\ \hline
		TVA & Trip-vehicle assignment formulation \\ \hline 
		RV graph & Request-vehicle graph \\ \hline
		RTV graph & Request-trip-vehicle graph \\ \hline
		ESV graph & Exclusive-sharing-vehicle graph \\ \midrule
		\multicolumn{2}{l}{Decision variables}                                                         \\ \midrule
	 	$\pi_{a}$ &    Trip-vehicle assignment policy                         \\ \hline
		$\pi_p$ &     Pricing policy                                          \\ \hline
		$p_e^r$ & MoD's price for request $r$ to use the exclusive service           \\ \hline
		$p_s^r$ & MoD's price for request $r$ to use the shared service     \\ \hline
		$y_i$ &  Binary variable for choosing matching $i\in \mathcal{M}$                                           \\ \hline
		$x_j$&   Overbooking penalty for vehicle $j \in \mathcal{V}$                                            \\ \midrule
		\multicolumn{2}{l}{Mixed MoD fleet parameters}                                                         \\ \midrule
		$\mathcal{R}$ & Set of trip requests  \\ \hline
		$\Phi_r$ &   Profit collected from request $r$                                     \\ \hline
		$Pr_e$ &   Probability of choosing the exclusive MoD service                                           \\ \hline
		$Pr_s$ &   Probability of choosing the shared MoD service                                           \\ \hline
		$Pr_o$ &   Probability of choosing the outside option                                         \\ \hline
		$c_e$ &  Cost of offering exclusive MoD service                     \\ \hline
		$c_s$ &  Cost of offering shared MoD service                        \\ \hline
		$c_{m, \text{operational} } $ & Operational cost for mode $m\in \{e,s\}$  \\  \hline 
		$c_{m, \text{potential} } $ & retrospective cost for mode $m\in \{e,s\}$  \\  \hline 
		$\Omega$ &  Maximum waiting time for matching                                             \\ \hline
		$\Delta $ &    Maximum delay                                         \\ \hline
		$U_m$ &    Customer's utility from choosing the mode $m \in \{e,s,o \}$                                          \\ \hline
		$w_m$ & Waiting time of mode $m$, $m\in \{e,s\}$ \\ \hline 
		$t_m$ & Travel time of mode $m$,  $m\in \{e,s\}$ \\ \hline 
		$\beta_{p}$ & Coefficient of price in utility function \\ \hline 
		$\beta_{w}$ & Coefficient of waiting time in utility function \\ \hline 
		$\beta_{t}$ & Coefficient of travel time  in utility function \\ \hline 
		$\mathcal{V}_e$ & Set of exclusive MoD vehicles \\ \hline 
		$\mathcal{V}_s$ & Set of shared MoD vehicles \\ \hline 
		$c_0$ & Operational cost if a request is not matched with any other request \\ \hline 
		$\alpha_{r'}$ & The probability request $r'$ is matched with vehicle carrying $r$ \\ \hline 
		$L$ & Total number of MoD vehicles in the cluster \\ \hline 
		$L_e$ & Total number of exclusive vehicles \\ \hline 
		$L_s$ & Total number of sharing vehicles \\ \hline 
		$Y$ & The steady-state throughput of MoD vehicles   \\ \hline
		$Y_e$ & The steady-state throughput of exclusive service vehicles  \\ \hline
		$Y_s$ & The steady-state throughput of shared service vehicles   \\ \hline
		$O_e$ & The number of available exclusive service vehicles \\ \hline 
		$O_s$ & The number of available shared service vehicles \\ \hline 
		$T_e$ & Average trip duration of exclusive services \\ \hline 
		$T_s$ & Average trip duration of shared services \\ \hline 
		$\eta_e$ & Average waiting time for exclusive service \\ \hline 
		$\eta_s$ & Average waiting time for shared service \\ \hline 
		$\zeta_s$ & Average utilization of sharing  service \\ \hline 
		$\bar{\epsilon}_o$ & Average profit if vehicle stays in the origin \\ \hline 
		$\bar{\epsilon}_d$ & Average profit if vehicle travels to destination \\ \hline 
		$\mathcal{M}$ & Set of ESV matching \\ \hline 
		$\mathcal{V}$ & Set of vehicles in the batch \\ \hline 
		$\mathcal{I}_{j }^\mathcal{M}$ & A set of matchings contains vehicle $j$ for given ESV matchings $\mathcal{M}$  \\ \hline 
		$\mathcal{I}_{k }^\mathcal{M}$ & A set of matchings contains requests $k$ for given ESV matchings $\mathcal{M}$  \\ \hline 
		$u_i$ & Expected profit of the ESV matching $i \in \mathcal{M}$ \\ \hline 
		$\gamma_{i,j}$ & Probability that vehicle $j$ is used if the ESV matching $i$ is selected \\  \hline  
		$\Phi_{r_1, r_2}$ & Expected profit of serving two requests $r_1, r_2$ in an ESV matching \\  \bottomrule
	\end{longtable}
	\end{center}

\section{Utilization of ridepooling vehicles in sequential pricing model} \label{sec:app2}
We use an aggregate mean utilization  $\zeta_s$ in the fleet conservation model. 
\begin{align}
	& Y_s(\eta, O) \triangleq \frac{ \zeta_s ( L_s - O_s)}{F_s(O_s) + T_s}.
\end{align}

We can derive this parameter from a Markov chain model of the ridepooling process.
The following formulation is for the special case that each ride takes at most two requests, which can be easily extended to higher capacity cases.
The number of customers on vehicle has three states $\{0,1,2\}$. The transition probabilities are $P_{ij}$ for $i,j \in \{0,1,2\}$. We let the parameters of  empty vehicle be $O_s, Y_s$  and vehicle with one customer be $O_s', Y_s'$.  It 

At the steady state, the number of vehicles in each state denoted by $N_i$ are:
\begin{align*}
	\begin{cases}
		N_0 = O_s + \eta_s Y_s \\
		N_1 = O_s' + \eta_s' Y_s' + T_s Y_s \\
		 N_2 = T_s' Y_s'
	\end{cases}.
\end{align*}

The detailed balance equations are:
\begin{align*}
	& N_0 P_{01} = N_1 P_{10} \\
	& N_1 P_{12} = N_ 2 P_{21}.
\end{align*}
We can use these equations to calibrate the throughput $Y_s, Y_s'$. For example, if we assume the that vehicles with zero and one customer are equal, we have $P_{01} = P_{12 }$. The average utilization is thus $\zeta_s = ( N_1 +2  N_2) / L_s $.   Note these two probabilities are roughly the ratio of time 
\begin{align*}
	\zeta_s = \frac{ O_s' + \eta_s' Y_s' + T_s Y_s + 2 T_s' Y_s' }{ L_s }
\end{align*}

\section{Proof of proposition~\ref{prop: seq}}\label{proof:prop:seq}
\begin{proof}
The partial derivative for $p_s$ is: 
\begin{equation}
        \frac{\partial \mathop{\mathbb{E}_{\pi_p}} [\Phi_r]}{\partial p_{s}} \\= \frac{\left(\beta \cdot e^{U_{s}} \cdot (p_{s} - m) + e^{U_{s}} \right) \cdot (e^{U_{s}} + e^{U_{e}} + D) - \beta \cdot e^{U_{s}} \cdot \left(e^{U_{s}}\cdot (p_{s} - m)+ e^{U_{e}} \cdot (p_{e} - c)\right)}{(e^{U_{s}} + e^{U_{e}} + D)^2}
\end{equation}

The partial derivatives are well defined everywhere. 
Since critical points are either points such that the partial derivatives do not exist or the partial derivatives are 0. 
For simplicity, let $e^{U_{s}}$ and $e^{U_{e}}$ be $x$ and $y$. Then, we can infer that $p_{s} = \frac{\log(x) - d_{s}}{\beta}$ and that $p_{e} = \frac{\log(y) - d_{e}}{\beta}$. A critical point will satisfy the following equations if it exists:
\begin{align}
     &x + y + y \cdot \log(\frac{x}{y}) - (d_{s} + \beta \cdot m - d_{e} - \beta \cdot c) \cdot y + D \cdot \log(x) + D \cdot (1 - d_{s} - \beta \cdot m) = 0 \label{equ: 1}, \\
      &x + y - x \cdot \log(\frac{x}{y}) + (d_{s} + \beta \cdot m - d_{e} - \beta \cdot c) \cdot x + D \cdot \log(y) + D \cdot (1 - d_{e} - \beta \cdot c) = 0 \label{equ: 2}
\end{align}

Let Equation~\eqref{equ: 1} subtract Equation~\eqref{equ: 2}, and we can find the following solution: 
\begin{equation}
    x = e^{d_{s} + \beta \cdot m - d_{e} - \beta \cdot c} \cdot y
\end{equation}

We can further infer that a critical point satisfies the following condition: 
\begin{equation}
    p_{e} - p_{s} = c - m
\end{equation}

Employing this relationship in Eq.~\eqref{equ: 1}, we can get the critical point: 
\begin{equation}
    p_e = \dfrac{\log\left(\dfrac{W\left((1 + e^{d_{s} - d_{e} + \beta \cdot m - \beta \cdot c}) \cdot e^{d_{e} + \beta \cdot c - 1}/ D\right) \cdot D)}{1 + e^{d_{s} - d_{e} + \beta \cdot m - \beta \cdot c}}\right) - d_{e}}{\beta}
\end{equation}
\end{proof}

\section{Proof of proposition~\ref{prop: bat}}\label{proof: prop: bat}
\begin{proof}
We first replace the price variables in the function by the following equations: 

\begin{equation}
    p_{1, s} = \dfrac{\log{\frac{D_1  P_{1, s}}{1 - P_{1, s} - P_{1, e}}} - d_{1,s}}{\beta}
\end{equation}
where $D_1$ is exponential to the power of the utility of $r_1$ taking the taxi service. Similarly, we can also use the predicted probabilities to represent the other price variables. Since the price vector and the predicted vector has a one-to-one matching~\parencite{li2011pricing}, we can obtain the price variables if we can find the optimal predicted probabilities. Therefore, minimizing the following function will be equivalent to maximizing the expected profit shown in Eq.~\eqref{equ: batchprofit}. \par

\begin{align}\label{equ: replaceobj}
    \beta  \mathbb{E}_{\pi_p} [\Phi_r]  &= P_{1, s}  \left(\log (\frac{D_1 \cdot P_{1, s}}{1 - P_{1, s} - P_{1, e}}) - \beta  c_{1, s} \right) 
   + P_{1, e}  \left(\log(\frac{D_1  P_{1, e}}{1 - P_{1, s} - P_{1, e}}) - \beta  c_{1, e}\right)  \nonumber  \\
   &+ P_{2, s}  \left(\log (\frac{D_2  P_{2, s}}{1 - P_{2, s} - P_{2, e}}) - \beta  c_{2, s}\right) 
   + P_{2, e}  \left(\log(\frac{D_2  P_{2, e}}{1 - P_{2, s} - P_{2, e}}) - \beta  c_{2, e}\right) \nonumber  \\
    &+ P_{1, s}  P_{2, s}  \beta  (c_{1, s} + c_{2, s} - c_{s, s})  
\end{align}

Let $\phi_1 := 1 - P_{1, s} - P_{1, e}$ and $\phi_2 := 1 - P_{2, s} - P_{2, e} $.  
The Hessian matrix of the above equation $H$ is as follows, where the order of each row and column is $P_{1, s}$, $P_{1, e}$, $P_{2, s}$ and $P_{2, e}$: 

\begin{align*}
    \begin{bmatrix}
    \dfrac{1}{P_{1, s}} + \dfrac{1}{\phi_1} + \dfrac{1}{\phi_1^2}       & \dfrac{1}{\phi_1} + \dfrac{1}{\phi_1^2} & \beta  (c_{1, s} + c_{2, s} - c_{s, s}) & 0 \\
    \dfrac{1}{\phi_1} + \dfrac{1}{\phi_1^2}       &  \dfrac{1}{P_{1, e}} + \dfrac{1}{\phi_1} + \dfrac{1}{\phi_1^2}  & 0 & 0 \\
    \beta (c_{1, s} + c_{2, s} - c_{s, s})  & 0 & \dfrac{1}{P_{2, s}} + \dfrac{1}{\phi_2} + \dfrac{1}{\phi_2^2} & \dfrac{1}{\phi_2} + \dfrac{1}{\phi_2^2} \\
    0       & 0 & \dfrac{1}{\phi_2} + \dfrac{1}{\phi_2^2} & \dfrac{1}{P_{2, e}} + \dfrac{1}{\phi_2} + \dfrac{1}{\phi_2^2}
\end{bmatrix}
\end{align*}

If Eq.~\eqref{equ: replaceobj} is jointly convex in the predicted probabilities, we know that the problem has at most one critical point.
We can find the optimal solution by finding the point satisfying the first-order derivative condition and compare its objective function value with the boundary points or using the second-order derivative test. 
If the above Hessian matrix is positive definite, we conclude that the convexity condition is satisfied. \par

Let $\Vec{y} = \{y_1, y_2, y_3, y_4\}$ be a non-zero vector of four any real numbers. We can derive the following: \par

\begin{equation}\label{equ: secondorder}
\begin{split}
    \Vec{y} \cdot H \cdot \Vec{y}^{T} &= (\frac{1}{1 - P_{1, s} - P_{1, e}} + \frac{1}{(1 - P_{1, s} - P_{1, e})^2})  (y_1 + y_2)^2 \\ 
    &+ (\frac{1}{1 - P_{2, s} - P_{2, e}} + \frac{1}{(1 - P_{2, s} - P_{2, e})^2})  (y_3 + y_4)^2 \\ 
    &+ \frac{y_1^2}{P_{1, s}} +  \frac{y_2^2}{P_{1, e}} + \frac{y_3^2}{P_{2, s}} +  \frac{y_4^2}{P_{2, e}} + 2 y_1 y_3 \beta  (c_{1, s} + c_{2, s} - c_{s, s})
\end{split}
\end{equation}

If $\frac{y_1^2}{P_{1, s}} + \frac{y_3^2}{P_{2, s}} + 2 y_1 y_3 \beta  (c_{1, s} + c_{2, s} - c_{s, s}) \geq 0$, $\Vec{y} \cdot H \cdot \Vec{y}^{T} \geq 0$ because all the other parts in the equation are non-negative. Let $C = c_{1, s} + c_{2, s} - c_{s, s}$ for readability, we can derive the following: 
\begin{equation*}
    \frac{y_1^2}{P_{1, s}} + \frac{y_3^2}{P_{2, s}} + 2 y_1 y_3 \beta  C = (\sqrt{-\beta  C}  (y_1 - y_3))^2 + (\frac{1}{P_{1, s}} + \beta  C)  y_1^2 + (\frac{1}{P_{2, s}} + \beta C)  y_3^2
\end{equation*}

Therefore, the condition to prove the problem is jointly concave in the predicted probabilities is the following: 
\begin{equation}\label{equ: oricondition}
    C \leq \min\{-\frac{1}{\beta  P_{1, s}}, -\frac{1}{\beta  P_{2, s}}\}
\end{equation}

Since $\beta < 0$, $P_{1, s} \leq 1$ and $P_{2, s} \leq 1$, the following condition is a sufficient condition to satisfy Eq.~\eqref{equ: oricondition}: 
\begin{equation}\label{equ: sufcondition}
    C \leq -\frac{1}{\beta}
\end{equation}
where $C = c_{1, s} + c_{2, s} - c_{s, s}$ is the operational cost savings if $r_1$ and $r_2$ can share a ride compared to the sum of the expected cost for either of them choosing the shared service. 

The cost savings should always be larger than 0. 
Otherwise, the service provider should not match $r_1$ and $r_2$ as a sharing trip. 
We know that $c_{s, s} \geq \max\{c_{1, e}, c_{2, e}\}$, $c_{1, s} \leq c_{1, e}$ and $c_{2, s} \leq c_{2, e}$, and thus Eq.~\eqref{equ: sufcondition} is satisfied if the following equation is satisfied: 
\begin{equation*}
    \min\{c_{1, e}, c_{2, e}\} \leq -\frac{1}{\beta}
\end{equation*}
\end{proof}

\end{appendices}

\printbibliography

\end{document}